\documentstyle[aps,prl,epsfig,floats]{revtex}   

\newif\ifpdf
\ifx\pdfoutput\undefined
  \pdffalse
\else
  \pdfoutput=1
  \pdftrue
\fi

\begin{document}
\draft
\twocolumn[\hsize\textwidth\columnwidth\hsize\csname@twocolumnfalse%
\endcsname 
\title{The ``square kagome'' quantum antiferromagnet and the eight vertex model}
\author{Rahul Siddharthan and Antoine Georges}
\address{
Laboratoire de Physique Th\'eorique, Ecole Normale
Sup\'erieure, 
24 rue Lhomond, 75231 Paris Cedex 05, France \\ and
Laboratoire de Physique des Solides, Universit\'e Paris-Sud,
B\^{a}t.~510, 91405 Orsay, France}

\date{March 12, 2001}
\maketitle

\begin{abstract}
We introduce a two dimensional network of corner-sharing triangles
with square lattice symmetry.  Properties of magnetic
systems here should be similar to those on the kagome lattice.
Focusing on the spin half Heisenberg quantum antiferromagnet, we
generalise the spin symmetry group from SU(2) to SU($N$).  In the
large $N$ limit, we map the model exactly to the eight vertex model,
solved by Baxter.  We predict an exponential number of low-lying
singlet states, a triplet gap, and a two-peak specific heat.  In
addition, the large $N$ limit suggests a finite temperature phase
transition into a phase with ordered ``resonance loops'' and broken
translational symmetry. 
\end{abstract}
]
\vspace*{0.5ex}

Frustrated magnetic systems have been attracting a lot of
attention in recent years.  One of the more interesting examples
is the spin half Heisenberg quantum antiferromagnet (QAF) on the
kagome lattice, a two dimensional network of corner-sharing
triangles with hexagonal voids.  From numerical
studies~\cite{lhuil} it is known that this model has a gap to
magnetic excitations, but this gap is filled with a continuum of
singlet excitations. The number of these excitations is estimated
to scale exponentially in the number of lattice sites $N_s$, as
$1.15^{N_s}$, and therefore there is a significant low temperature
entropy in the thermodynamic limit. It is believed that the
low-temperature physics would be well described by a
resonating valence bond (RVB) picture.

Some insight into this has been gained by generalising the
symmetry group to SU($N$) and going to the large $N$ limit with a
particle-hole symmetric fermion representation of the spins.  It
was proved by Rokhsar~\cite{rokhsar} that for most common
lattices, any ``fully dimerised'' state (in which every site is
part of a dimer pair with another site) is a ground-state of the
$N=\infty$ model. The ground-state is thus macroscopically
degenerate in this limit ~\cite{sachdev}. Marston and
Zeng~\cite{marston} applied this picture to the kagome lattice.
For the physical SU(2) system, superpositions of all such states,
with dimers being interpreted as singlet pairings between the
respective spins, would be good candidates for the low-lying
singlet states. However, it is clear that a further selection of
states occurs when going from $N=\infty$ to $N=2$ since the number
of dimer coverings rises with system size as $1.26^{N_s}$~\cite{elser} 
rather than the observed $1.15^{N_s}$. 

\begin{figure}[ht]
\begin{center}
 \ifpdf
  \epsfig{file=sqkag.pdf, width=6cm, clip=}
 \else 
  \epsfig{file=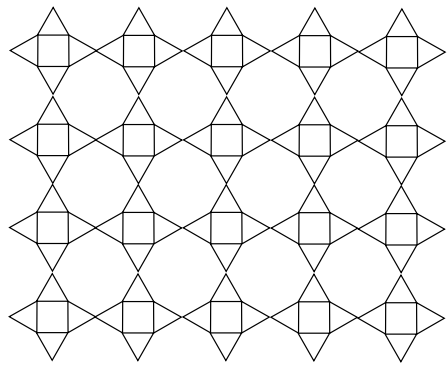, width=6cm, clip=}
 \fi
\end{center}
\caption{\label{sqkag}
 The ``squagome'' lattice introduced in this paper.
 }
\end{figure}

In this paper we introduce a lattice~\cite{senconv}
(figure~\ref{sqkag}) with square lattice symmetry, on which, we
believe, magnetic properties should be similar to those of the
kagome lattice. This, too, is a two dimensional network of corner
sharing triangles, but the voids in between are alternately
squares and octagons, rather than hexagons. We therefore
name it the ``square kagome'' (or ``squagome'') lattice. Using
the large-$N$ limit as a guidance, we are able to make precise
statements on the Heisenberg QAF on this lattice. At $N=\infty$,
the ground-state is again exponentially degenerate, corresponding
to dimer coverings. We demonstrate that, to next order in the
$1/N$ expansion, an {\it exact mapping} can be made to the
classical eight-vertex model on the square lattice, with an
additional two-fold degeneracy per vertex. As a result, a
finite-temperature phase transition is found corresponding to the
breaking of a discrete symmetry and to the dominance of specific
dimer patterns in the low-temperature phase. The ground-state
degeneracy is partially lifted to this order, leading to
exponentially many excited singlet states below the triplet gap. Further
lifting of the ground-state degeneracy is expected to occur at
higher orders in the $1/N$ expansion. We fully expect that at
least some of these features will persist in the SU($2$) model,
which will thus have an exponential number of low-lying singlet excitations,
a triplet gap, and RVB-like low temperature states, just as in the
kagome lattice. Whether a finite-temperature phase transition also
applies to the SU($2$) case is an intriguing possibility which
deserves further investigations.

As in \cite{readsach,marston} we use a particle-hole symmetric
fermionic representation of SU($N$) spins, corresponding to the
local constraint $ \sum_{\alpha} f^\dag_{i\alpha}f_{i\alpha}=N/2$
at each lattice site. The Hamiltonian reads:
\begin{equation}
H = \frac{J}{N} \sum_{\langle ij \rangle}
   f^\dag_{i\alpha}f_{i\alpha'}\,
  f^\dag_{j\alpha'}f_{j\alpha}
\end{equation}
where $\alpha$, $\alpha'$ range from 1 to $N$, and summations over
repeated indices are implied.
Introducing a Hubbard-Stratonovich field $Q_{ij}(\tau)$ on each
bond, conjugate to $\sum_\alpha f^\dag_{i\alpha}f_{j\alpha}$, and
implementing the constraint using a Lagrange multiplier
$\lambda_i(\tau)$ on each site leads to the following
imaginary-time effective action, after integrating out the
fermions \cite{readsach}:
\begin{eqnarray}\nonumber
S_{\rm eff}/N & = &\int_0^{\beta}d\tau\, [\frac{1}{J}\sum_{\langle
     ij\rangle} |Q_{ij}|^2\,-\,\sum_i\lambda_i \\ 
 & &-\mbox{Tr}\;\ln\left(\partial_\tau\delta_{ij}
          +\lambda_i\delta_{ij}+Q_{ij}\right)]
\end{eqnarray}
At $N=\infty$, one has to search for saddle-points of this
effective action. There are exponentially many saddle points with
the lowest energy (as in \cite{rokhsar,marston}), given by all
``dimer coverings'' in which every site is paired uniquely with
one of its nearest-neighbour and $Q_{ij}$ $=$ $Q$ ($=J/2$ at
$T=0$) on dimer bonds and zero otherwise. The $\lambda_i$'s are
zero at the saddle point. The physical interpretation of a dimer
is the formation of a singlet bond between the two sites.

\begin{figure}[ht]
\begin{center}
 \ifpdf
  \epsfig{file=eightvtx.pdf, width=8cm, clip=}
 \else
  \epsfig{file=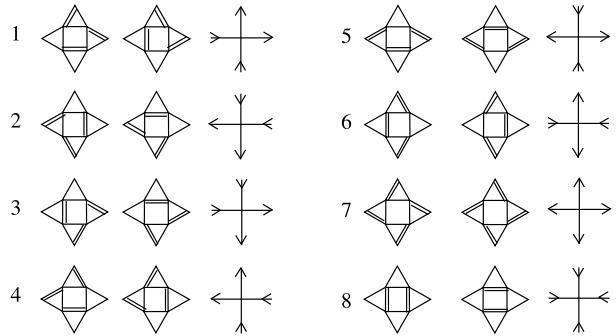, width=8cm, clip=}
 \fi
\end{center}
\caption{\label{eightvtx}
 The 16 configurations of an individual plaquette, which map onto
 the 8 allowed vertices in the eight vertex model, with a degeneracy
 of 2 per vertex.
 }
\end{figure}

When studying dimer coverings on the squagome lattice, it is
useful to look at individual plaquettes
of four triangles enclosing a square. The entire lattice can be
viewed as a network of such plaquettes joined at corners.  One can
convince oneself that if each corner of the internal square is to
be part of a dimer, then the number of external corners which are
parts of dimers in this plaquette will always be even.  Moreover,
in two such plaquettes joined at a corner, that corner must be
part of a dimer in one plaquette and not another plaquette.  A
consistent scheme for representing this is to draw an arrow
pointing out of the plaquette when that corner is part of a dimer,
and into the plaquette when the corner is not part of a dimer.  It
turns out that there are exactly sixteen allowed configurations
per plaquette, as illustrated in figure~\ref{eightvtx}.

The remarkable fact is that when we picture the system thus in
terms of arrows, {\em what we have is precisely the eight vertex
model on the square lattice} which was solved exactly by Baxter in
the 1970s, and discussed in detail in his book~\cite{baxter}.
There are, however, two possible dimer configurations per vertex,
which introduces an extra twofold degeneracy for the original
model. In the infinite~$N$ limit (zeroth order in the large $N$ expansion), 
all vertices have equal weight; the energy is of order $NJ$ and so
is the ``triplet gap''. 
We now consider the first order corrections to this infinite~$N$
picture, which we expect to lift the degeneracy between different
vertices.  Read and Sachdev~\cite{readsach} showed that
the first correction beyond $N=\infty$ lowers the
energy of configurations in which two dimers sit on the same
square plaquette.
This leads for example to a columnar dimer order on the square
lattice. In the  squagome lattice, we find that the same
reasoning leads to a lowering of the energy of vertex 8 compared
to the others (fig.\ \ref{eightvtx}). To reach this conclusion, we follow
\cite{readsach} and expand around a dimerised saddle-point:
$Q_{ij}=Q\,D_{ij}+\delta\,Q_{ij}$. $D_{ij}$ specifies the dimer
pattern ($=1$ if bond $(ij)$ has a dimer, $=0$ otherwise) and
$\delta\,Q_{ij}(\tau)$ is a fluctuation. Expanding the effective
action to quadratic order in the fluctuations, one finds two types
of terms: bond-diagonal terms involving
$Q^2D_{ij}^2\delta\,Q_{ij}^2$ and off-diagonal terms of the form:
$Q^2\delta\,Q_{ij} D_{jk} \delta\,Q_{kl} D_{li}$.
The latter can be non-zero only on a square-plaquette
configuration where a pair of opposite sides has dimer bonds and
the other two bonds have fluctuations. This first order correction
is of order $1$ in energy, and can be thought of as a
`resonance' of the two possible dimer configurations on a square.
Only the off-diagonal
contributions change the relative energies in our vertex model,
hence the lowering of vertex 8 associated with square patterns.

To second and higher order, too, the only off-diagonal
contributions come from loops.  Thus, the hexagonal dimer
configurations in vertices 1--6 will also have a lowering of
energy, of order $J/N$, as pointed out by 
Marston and Zeng~\cite{marston} for the kagome lattice.
The resonance of an octagonal loop in vertex 7 is at still higher
order. So both these can be ignored in the first order approximation.
In addition, second 
order corrections will lift the degeneracy between the two dimer
configurations on a square plaquette, favouring a resonating
combination. The comparison to the SU(2) case is instructive: for
a loop with an even number of sites, the two possible dimerised
states are not eigenstates, but superpositions of such states have
lower energy expectation values than the pure dimerised states;
the energy gain decreases exponentially with increasing loop
length. This is the idea behind the ``quantum dimer'' approach
\cite{rokhkivel,zengelser}. The $1/N$ expansion is another approach to an
expansion in the size of increasingly long resonance loops.

So we have an 8 vertex model with vertex 8 having an energy, of
say $-2\epsilon$,  and all other vertices having
zero energy. But with periodic boundary conditions, vertices 7 and
8 must occur in equal numbers (because they are respectively
``sources'' and ``sinks''); so it causes no error to assign them
equal energies of $-\epsilon$ each.  In this way we have mapped
our system to an eight vertex model, with weights of unity for all
vertices satisfying the ``ice rule'' and higher weights for the
remaining two vertices, and with an additional ``internal''
degeneracy of 2 per vertex.
Since vertices 7 and 8 occur in pairs and together contain two
defective triangles in eight, and all other vertices contain
exactly one defective triangle in four, the total number of
defective triangles is always $1/4$ the total number of
triangles---just as in the kagome case \cite{elser}.

We now draw some conclusions on the physics of the {\it squagome}
QAF, using the mapping on the eight vertex model. Three properties
{\it of the vertex model} play an important role in the following.
First, its ground state is two-fold degenerate, with the
configuration consisting of an alternation of vertices 7 and 8.
Second, there are no gapless excitations, but there is a minimum
gap of order $\epsilon$ between any two levels. Excited levels
are, in general, degenerate. Third, the vertex model undergoes a
phase transition from an ordered state to a disordered,
``paramagnetic'' state. One should bear in mind, however, that
the underlying dimer model has an additional 2-fold degeneracy
per plaquette (figure~\ref{eightvtx}).
So the ground state has, to this order, a
degeneracy of order $2^{N_p}$ with $N_p$ the number of plaquettes, 
and each excited level has at least
this degeneracy too. 
The spacing between levels is of order $\epsilon \propto J$, which
is much smaller than the triplet
gap which is of order $NJ$.
So already we have a picture of an exponential number
of singlet states below the triplet gap.  We expect
that the degeneracy $2^{N_p}$ of each excited level will be lifted
with further $1/N$ corrections.  For finite $N$, each level could
then broaden into a
band. 
The exponential ground-state
degeneracy (associated with the local twofold degeneracy present
at first order) will also be lifted at higher order. 
So in the
SU(2) case, our picture based on this simple mapping is that of a
system with a triplet gap and an exponential number of singlet
excitations of order $2^{N_p}$ $\approx$ $1.12^{N_s}$ since
$N_p=N_s/6$. This compares well with the commonly accepted picture
of the Heisenberg antiferromagnet on the kagome lattice, where the
number of singlet states is of the order of $1.15^{N_s}$.

A somewhat different physical picture for this number was recently
proposed by Mambrini and Mila~\cite{mila2}, who
take as a starting point a lattice of decoupled up-pointing triangles
(with internal coupling $J_1$), and show that the degenerate ground
state broadens into a band as the inter-triangle coupling $J_2$ is
turned on. However, in any dimer covering of the kagome lattice,
exactly $1/4$ of the triangles are left without any dimerised side,
and it is not clear that when $J_2$ equals $J_1$, it is a good
approximation to assume that all these ``defective'' triangles belong
to one sub-lattice---even though that assumption yields the desired
number of singlet states.  Work on simpler models~\cite{sen} suggests
that, quantitatively, this approximation is not in fact very accurate.
The approach has its merits, however, and can also be applied to the
squagome lattice; the results, we believe, will be similar.

\begin{figure}[ht]
\begin{center}
 \ifpdf
  \epsfig{file=spcorr.pdf, width=5cm, angle=270, clip=} 
 \else
  \epsfig{file=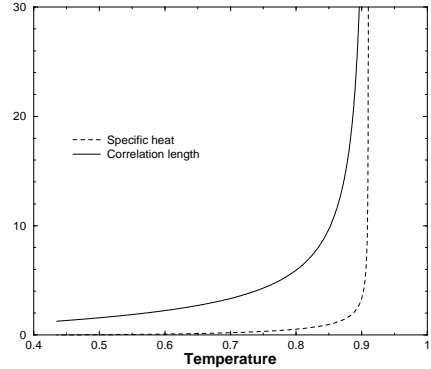, width=5cm, angle=270, clip=} 
 \fi
\end{center}
\caption{\label{spcorr}
 The specific heat, and correlation length, of the eight vertex model approximation
 to the squagome lattice (temperature in units of $\epsilon$).
 }
\end{figure}

Using our mapping and Baxter's results~\cite{baxter}, we can
approximate the thermodynamics of the squagome QAF at low
temperatures (below the triplet gap).  Fig.~\ref{spcorr} displays the
calculated correlation length and specific heat as a function of
temperature. The latter displays a sharp peak near the transition,
which takes place on a scale $\epsilon$ much smaller than the
triplet gap. This is reminiscent of the lowest specific-heat peak
reported in quantum dimer model based studies of the kagome antiferromagnet
\cite{zengelser}. Since the triplet excitations will also contribute
a peak at higher energy, we see that this system, too, shows the two-peak 
feature which was noticed early in the kagome QAF~\cite{elser}.

\begin{figure}[ht]
\begin{center}
 \ifpdf
  \epsfig{file=eldens.pdf, width=4.5cm, clip=}
 \else
  \epsfig{file=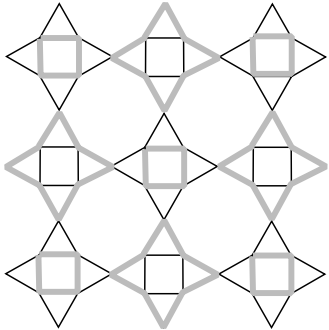, width=4.5cm, clip=}
 \fi
\end{center}
\caption{\label{eldens}
  The ground state ordering seen in the large $N$ limit may possibly
 manifest itself in the physical system by increased electron density
 along the thick grey lines here.
 }
\end{figure}

An intriguing feature of our results is a hidden ordering in the
ground state. At first order beyond $N=\infty$, this ordering
corresponds to a staggered pattern in which every other plaquette
is in one of the two configurations corresponding to vertex 7, and
its neighbors in one of the two configurations corresponding to
vertex 8.
At higher order, the picture of two equal-energy configurations
per vertex will not persist: such configurations will in general
mix, leading to a splitting of energies. If the ordering still exists, 
it would then correspond to a staggered ordering of
plaquettes in which the resonating dimers live on the squares on
every other plaquette, and on the star-shaped boundary on the
neighbouring plaquettes (figure~\ref{eldens}).  Such an ordering
in a real system may manifest itself as an additional
electron density along the resonant squares (and octagon stars),
much as happens with the hexagonal ring in benzene. This 
may be detectable via STM experiments.  But it is not clear whether
this ordering would actually persist in the SU(2) system, or be washed out
by further $1/N$ corrections.  
Since the ordering consists of alternate
plaquettes having dimer pairs in their central squares, a possible
order parameter could be the quantity $({\bf S}_1$ $ +$ $ {\bf
S}_2$ $ +$ $ {\bf S}_3 $ $+$ $ {\bf S}_4)^2$ where these are the
four spins on the square; this is minimised when two opposite
sides are paired as singlets.  This is not the true SU(2) ground
state, but it is possible that such states will dominate a true
RVB-like ground state.  It is also not clear whether a sharp phase
transition will persist in the SU(2) case, but we note that this
ordering is not ruled out at finite temperatures by the
Mermin-Wagner theorem since it
originates from the breaking of a {\em discrete} translational
symmetry. So the intriguing possibility of such a phase transition
in a 2D Heisenberg system exists.

\begin{figure}[ht]
\begin{center}
 \ifpdf
  \epsfig{file=kagplaq.pdf, width=4.5cm, clip=}
 \else
  \epsfig{file=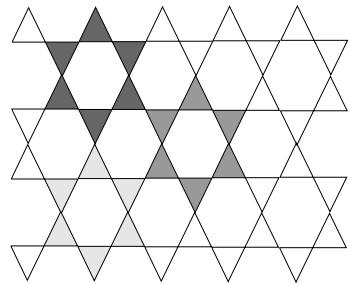, width=4.5cm, clip=}
 \fi
\end{center}
\caption{\label{kagplaq}
 The kagome lattice divided into star-shaped plaquettes which
 form a triangular lattice, by analogy with our treatment of the
 squagome lattice.
 }
\end{figure}

The obvious question at this point is whether such a study can be
made of the kagome lattice too. This lattice can be decomposed
into star-shaped plaquettes of hexagons bordered by triangles,
which sit at the sites of a triangular lattice
(figure~\ref{kagplaq}). In each such plaquette, again, the
requirement that each internal site must be part of a dimer pair
implies that of the six external sites, an even number must be
parts of dimers. But to progress beyond that is difficult, for
several reasons. First, the underlying lattice is a triangular
lattice, with a high coordination number. Second, each vertex has
six (rather than four) arms, and the even-number restriction still
leads us to 32 kinds of vertices---each being again two-fold
degenerate. There is thus no hope of an exact solution. Estimating
vertex weights is possible in principle but requires us in this case
to go to {\it second order} in the $1/N$ expansion. 
As noted before~\cite{marston}, hexagons with three dimerised sides will be
preferred. This only fixes the weight of one of the 32 vertices,
and since every vertex is now a source or a sink, it is impossible
to use this statement to fix the weight of any other vertex.
Nevertheless, if we assume that such ``perfect hexagons'' will
dominate, one should be able to maximise their number by forming a
regular lattice of them, and they can be detected by an order
parameter which is the total spin on the six sites of the hexagon,
in analogy to the square plaquette order parameter above. Note
that there is still a hidden degeneracy of 2 per plaquette which
gives rise to $2^{N_p}$ $\approx$ $1.08^{N_s}$ states, since in
this case the number of sites $N_s$ $=$ $9 N_p$.  The observed
number of low-energy singlets, $1.15^{N_s}$, suggests a
significant additional degeneracy from the number of allowed
vertex configurations.  So even in a large $N$ limit, the kagome
ground state may not be as highly ordered as the squagome.

If a hidden ordering does exist in the kagome case,
it may correspond to a pattern of hexagon-shaped resonances.
However, as described above, our other conclusions about the
squagome lattice are very well corroborated by what is known
about the kagome lattice, and in general we expect these systems to
behave very similarly.  

In conclusion, we have displayed a lattice, which we call the 
squagome lattice, which is conceptually very similar to the
kagome lattice, but with square-lattice symmetry.  We have argued
that physical properties of spin systems should be very similar on
this lattice to properties on the usual kagome lattice. We have
shown that, to next to leading order in a $1/N$ expansion, an
exact mapping exists between the SU($N$) QAF on this lattice and
the classical eight-vertex model. This allows to draw several
conclusions on the physics of this QAF at large-N, some of which
are likely to extend to the physical SU($2$) case. In particular,
we point out the intriguing possibility of
a finite-temperature long-range ordering of the resonance loops.
Perhaps most notably, we have connected the 
field of frustrated quantum systems to a classic exactly
solved problem of statistical mechanics.

We acknowledge useful discussions with C. Lhuillier, R. Moessner and
D. Sen.  The LPT-ENS is ``Unit\'e Mixte de Recherche CNRS UMR-8549,
associ\'ee \`a l'Ecole Normale Sup\'erieure''. The LPS-Orsay is
``Unit\'e Mixte de Recherche CNRS UMR-8501, associ\'ee \`a
l'Universit\'e Paris-Sud''.


\begin{thebibliography}{88}
\bibitem{lhuil} C. Waldtmann, H.-U. Everts, B. Bernu, C. Lhuillier, P.
Sindzingre, P. Lecheminant, L. Pierre, Eur.\ Phys.\ J. B {\bf 2}, 501
(1998)
\bibitem{rokhsar} D. S. Rokhsar, Phys.\ Rev.\ B {\bf 42}, 2526 (1990)
\bibitem{sachdev} S.~Sachdev [Phys.\ Rev.\ B {\bf 45}, 12377 (1992)]
considered bosonic representations of the symplectic group Sp($N$)
on the kagome lattice, which lead to the quite different picture
of a unique RVB ground-state with a gap, or to long-range order at
higher values of the spin. Fermionic representations of Sp($N$) do
lead to a similar dimerised picture as for SU($N$).
\bibitem{marston} J. B. Marston and C. Zeng, J. Appl.\ Phys.\ {\bf
69}, 5962 (1991)
\bibitem{elser} V.~Elser, Phys.\ Rev.\ Lett.\ {\bf 62}, 2405 (1989)
\bibitem{senconv} The idea of such a lattice came up in a conversation
R.~S. had with Diptiman Sen.
\bibitem{readsach} N. Read and S. Sachdev, Nucl.\ Phys.\ B {\bf 316},
609 (1989)
\bibitem{baxter} R.~J.~Baxter, {\em Exactly Solved Models in Statistical
  Mechanics}, Academic Press (1982)
\bibitem{rokhkivel} D. Rokhsar and S. Kivelson, Phys.\ Rev.\ Lett.\ {\bf 61},
2376 (1988)
\bibitem{zengelser} C. Zeng and V. Elser,  Phys.\ Rev.\ B {\bf
51}, 8318 (1995).
\bibitem{mila2} M.~Mambrini and F.~Mila, Eur.\ Phys.\ J. B {\bf 17}, 651 (2000)
\bibitem{sen} C.~Raghu, I.~Rudra, S.~Ramasesha, D.~Sen, Phys.\ Rev.\
 B {\bf 62}, 9484 (2000)
\end{thebibliography}
\end{document}